\documentclass[manuscript,screen,review]{acmart}

\setcopyright{rightsretained}
\acmJournal{TOMS}
\acmYear{2026}
\acmVolume{1}
\acmNumber{1}
\acmArticle{1}
\acmMonth{1}
\acmDOI{10.1145/XXXXXXX}

\usepackage{listings}
\usepackage{siunitx}

\lstdefinestyle{tightcode}{
  language=C,
  basicstyle=\ttfamily\small,
  columns=fixed,
  basewidth=0.55em,   % horizontal spacing between characters
  lineskip=-2pt,      % vertical spacing between lines
  aboveskip=1em,
  backgroundcolor=\color{black!8},
  frame=tlrb,
  framerule=0pt,
  framesep=5pt,
  xleftmargin=4em,
  xrightmargin=4em
}

\newcommand{\NA}{\multicolumn{1}{c}{--}}

\AtBeginDocument{%
  }

\begin{document}

%%
%% Title
%%
\title{Randompack: Cross-Platform Reproducible Random Number
Generation and Distribution Sampling}

%%
%% Author information - FILL IN YOUR DETAILS
%%
\author{Kristján Jónasson}
\authornote{Corresponding author.}
\email{jonasson@hi.is}
\orcid{0000-0000-0000-0000}
\affiliation{%
  \institution{University of Iceland}
  \city{Reykjavík}
  \country{Iceland}
}

%%
%% Abstract
%%
\begin{abstract}
  A C library for random number generation, Randompack, is presented.
  The library implements several modern random number generators
  (engines), including xoshiro256, PCG64, Philox, ranlux++, and sfc64;
  14 continuous distributions including uniform, normal, exponential,
  gamma, beta, and multivariate normal; raw bit streams, bounded
  integers, permutations, and sampling without replacement. The engine
  and the distribution layers are separated so any engine can be used
  with any distribution. Benchmarks show that Randompack is faster
  overall than competing libraries, with speedup factors ranging from
  about 1 to 15 depending on engine, distribution, interface, and
  platform. A distinguishing feature is reproducibility: with the same
  seeds Randompack gives compatible results across programming
  languages, computers, CPU architectures, and compilers. The library
  includes comprehensive support for parallel simulation. It is
  accompanied by a comprehensive test suite, benchmarking programs,
  and example programs. Interfaces to Fortran, Python, Julia, and R
  have been implemented; their benchmark results are included,
  although their design and implementation are otherwise outside the
  scope of the article. Unlike other available C libraries with
  comparable scope, Randompack is permissively licensed under the MIT
  license, and it is open source and publicly available through GitHub
  and conda-forge.
  
\end{abstract}

%%
%% ACM Computing Classification System
%%

%%
%% Keywords
%%

\maketitle

\section{Introduction}

The computer generation of pseudo-random numbers has a history dating
back to the first computers, with von Neumann's middle-square idea
(1946) \cite{vonneumann1951rejection,lecuyer2017history} and Lehmer's
linear congruential generator (LCG, 1949)
\cite{lehmer1951mathematical,lecuyer2017history}. These first random
number generators (RNGs) were aimed at generating random integers
(typically in $[0, 2^k)$) and uniformly distributed reals in $[0,1]$
(through division by $2^k$). Other important 20th century milestones
of this type are the Tausworthe RNGs (1965)
\cite{tausworthe1965linear}, the Wichmann-Hill RNG (1982)
\cite{wichmann1982efficient}, the Park-Miller LCG (1988)
\cite{park1988random}, RANLUX (1994) \cite{luscher1994portable},
adopted in the CERN ROOT framework, the Mersenne Twister (1998)
\cite{matsumoto1998mersenne} with its extremely long period,
$2^{19937}-1$, and, in 1999, MRG32k3a \cite{lecuyer1999good} to be
discussed further below.  

In parallel with the development of uniform generators, methods for
generating non-uniform variates were developed. Von Neumann's 1951
paper \cite{vonneumann1951rejection} had a clever algorithm for
drawing exponentially distributed samples, also obtainable by taking
the logarithm of uniform samples. A seminal method for random normals
was given by Box and Muller in 1958 \cite{box1958normal}. The normal
and the exponential distributions play an important role. The gamma
distribution is another important case. Random sampling from several
other distributions can be done by combining samples from these three
(see Section \ref{sec:cont-distr}). The Box-Muller method requires the
evaluation of a square root, a logarithm, a sine, and a cosine for
each pair of samples, but a few years later Marsaglia and Bray
proposed the polar method where the two trigonometric functions were
replaced by rejection sampling with acceptance rate $\pi/4$. Rejection
sampling had already been used by von Neumann in his exponential
method \cite{vonneumann1951rejection}. It also plays a central role in
the ratio of uniforms method (1977)
\cite{kinderman1977ratio,leva1992fast} which has an acceptance rate of
73\% and needs two uniform samples but only 0.012 logarithms and no
square roots per normal sample. The next advance, also using
rejection, was the ziggurat method (1984, 2000) proposed in two stages
by Marsaglia and Tsang \cite{marsaglia1984fast,marsaglia2000ziggurat},
the revised method drawing from the tail using an earlier idea
\cite{marsaglia1964tail}. The seminal article from 2000 gives ziggurat
algorithms for both normal and exponential variates. In the same year
Marsaglia and Tsang published a notably elegant method for the gamma
distribution, later adopted by many RNG libraries
\cite{marsaglia2000gamma}. Earlier gamma algorithms were mostly
rejection-based and often quite complex.

Around the turn of the century, random number research placed
increasing emphasis on the statistical quality of RNGs, on devising
methods for creating independent streams of random numbers, and on
ensuring long periods, spurred by faster, often parallel computers.
For example, on a modern computer, the output of Park–Miller and other
32-bit-state RNGs repeats itself after only a few seconds on a single
core of a modern computer. This development increased the importance
of empirical testing. Knuth had already described rigorous statistical
tests of RNGs in 1969 in \emph{The Art of Computer Programming}
\cite{knuth1969art}, but the next major milestones were the Diehard
suite in 1996 \cite{marsaglia1996diehard}, DieHarder in 2006
\cite{Brown2006DieHarder}, TestU01 in 2007 \cite{lecuyer2007testu01}
with its well-known SmallCrush, Crush, and BigCrush batteries, and
PractRand \cite{dotyhumphrey2010pract}. TestU01 exposed weaknesses
in many widely used RNGs, which failed its crush-batteries. More
recently, Foreman et al. \cite{foreman2024statistical} described an
openly available RNG testing environment built on existing suites. In
parallel with these developments, standards agencies on both sides of
the Atlantic developed methods for testing cryptographic RNGs
\cite{schindler2002evaluation,rukhin2010nist,saarinen2022sp}.

We now return to the development of RNG engines, continuing from the
widely used Mersenne Twister (MT; Table~\ref{tab:language-rngs}).
One of the first RNGs to offer independent streams, paving the way for
the subsequent advances was the aforementioned MRG32k3a, with a period
of $2^{191}$, introduced by L'Ecuyer in \cite{lecuyer1999good} and
further developed for parallel use in 2002 \cite{lecuyer2002streams}.
The WELL generators \cite{panneton2006well} were intended to (and did)
improve on the MT statistical quality although they did not obtain
widespread adoption in software. By contrast, the RNGs sometimes called 
the xor family, pioneered by Marsaglia in 2003 with
xorshift and XORWOW \cite{marsaglia2003xorshift}, and further
developed, primarily by Vigna, in a series of articles
\citep{panneton2005xorshift,vigna2016experimental,vigna2017further,
  blackman2021scrambled}. The current flagships are the closely
related xoshiro256++ and xoshiro256**, that have been adopted as
default generators by Rust, Julia, C\#, and gfortran
(Table~\ref{tab:language-rngs}). All generators in the xor family
support so-called \emph{jumping}, whereby the state can be advanced
efficiently by very large amounts, such as $2^{128}$ steps, to create
independent streams. Such parallelization is sometimes referred to as
block-splitting. The aforementioned Practrand suite included an RNG,
sfc64 (\emph{small-fast-counting}), that also allows independent
streams. Interestingly, sfc64 is planned to become the default random
number generator in a forthcoming release of the Los Alamos MCNP
radiation-transport code.
\cite{josey2023reassessing,gibson2025mcnpnewsletterq1}.

Some important pseudorandom number generators (PRNGs) emerged from the
search for fast software stream ciphers. Such ciphers can also serve
as cryptographically secure PRNGs (CSPRNGs), notably ChaCha20 (2008)
\cite{bernstein2008chacha}. Another important milestone came in 2011
with the Random123 library by Salmon and Moraes
\cite{salmon2011parallel}. Its counter-based Threefry and Philox
generators have become popular for simulation on GPUs and are the
default in several systems (Table~\ref{tab:language-rngs}). They allow
the generation of a vast number of independent streams.

Another important development was O'Neill's PCG family (2014)
\cite{oneill2014pcg}. It also supports many streams via a
stream-selector termed \emph{increment}. A 64-bit version of it has
been adopted as NumPy's default RNG engine. In 2019 O'Neill suggested
an improved version, PCG64-DXSM, which is planned to become default
for NumPy \cite{oneill2019pcg64dxsm-post,numpy_pcg64dxsm_upgrade}. One
more way of getting many streams is via stream splitting, an
object-oriented approach implemented in SplitMix (2014)
\cite{steele2014splitmix} and refined in the LXM family (2021)
\cite{steele2021lxm}. These RNGs provide both \emph{generate} and
\emph{split} as primitive operations, the latter creating two
independent generators. Java's default RNG is from the LXM family,
while SplitMix has primarily found application in spreading an initial
seed across the state of other generators.

In 2017 L'Ecuyer published a comprehensive history of uniform
pseudo-random number generators \cite{lecuyer2017history}, which gives
a fuller treatment of the generators discussed above. He also covers
several generators not included here: RANDU, RANNO (both notoriously bad),
lagged Fibonacci, generalized feedback shift register (GFSR),
add-with-carry, subtract-with-borrow (which underlies RANLUX), and
block-cipher-based counter generators. The last category includes
AES-CTR, which helped pave the way for Random123. The earlier article
by James (1990) \cite{james1990review}, as well as the TestU01 article
\cite{lecuyer2007testu01}, also provide useful historical overviews.
More recent, though in both cases limited, reports on RNG testing are
given by \cite{bhattacharjee2022search} and
\cite{josey2023reassessing}.

\begin{table}
\caption{Default random number generators in selected languages and
  libraries. The generators are discussed briefly in the main text.
  ChaCha8 and ChaCha12 are exactly like ChaCha20 except that the core
  loop takes fewer rounds. The Mersenne Twister is in all cases the
  original 32-bit MT19937 version.}
\label{tab:language-rngs}
\centering
\begin{minipage}{0.52\columnwidth}
\centering
\begin{tabular}{l@{\hspace{2.5em}}l}
\toprule
Language/library          & Default engine \\
\midrule
Go                        & ChaCha8 \\
Rust (rand default)       & ChaCha12 \\
Java                      & LXM (L32X64MixRandom) \\
Boost                     & Mersenne Twister\textsuperscript{a} \\
C++ standard library      & Mersenne Twister\textsuperscript{a} \\
Intel oneMKL              & Mersenne Twister\textsuperscript{a} \\
GNU Scientific Library    & Mersenne Twister \\
Matlab (core)             & Mersenne Twister \\
Python random module      & Mersenne Twister \\
PyTorch (CPU)             & Mersenne Twister \\
Ruby                      & Mersenne Twister \\
R (base)                  & Mersenne Twister \\
R (parallel)              & MRG32k3a \\
R (dqrng)                 & xoroshiro (128++) \\
NumPy                     & PCG64 (XSL-RR 128/64) \\
OpenRAND                  & Philox \textsuperscript{a} \\
PyTorch (GPU)             & Philox \\
TensorFlow                & Philox \\
Google JAX                & Threefry \\
Matlab (parallel toolbox) & Threefry \\
CUDA cuRAND               & XORWOW \\
JavaScript (V8)           & xorshift (128+) \\
C\# (.NET)                & xoshiro (256**) \\
gfortran                  & xoshiro (256**) \\
Julia                     & xoshiro (256++) \\
Rust (rand SmallRng)      & xoshiro (256++) \\
\bottomrule
\end{tabular}
\par\medskip
\footnotesize
\raggedright
\textsuperscript{a}No unique default RNG exists; the table reports the
engine most commonly used in the official documentation and examples.
\end{minipage}
\end{table}

Randompack is a C library providing a selection of modern,
high-quality, and efficient pseudo-random number generators together
with a range of continuous and discrete distributions. It is intended
to fill a gap in the random-number-generation support available in
programming languages and libraries. The C libraries that come closest
are the GNU Scientific Library (GSL), UNURAN
\cite{hormann2000randomnumber}, the NAG library, and Intel oneMKL.
However, none of these libraries are permissively licensed, and only
the first two are open source. Moreover, those two are somewhat dated
and rely largely on older RNG engines. Another C library worth
mentioning is PRAND \cite{barash2014prand}, which supports GPUs and
CPUs, implements several engines including the Mersenne Twister and
MRG32k3a, and supports block-splitting, but only provides $U(0,1)$ and
raw bit streams and is non-permissively licensed.

C is a particularly suitable implementation language because it can be
interfaced readily from many other programming languages. This makes
it possible to aim not only at exact repeatability across computers,
as in some existing systems (e.g. NumPy), but also at exact
repeatability across programming languages. Randompack interfaces to
Fortran, Julia, Python, and R have been developed, but their design
and implementation are outside the scope of the present article, which
focuses on the C library itself (see Section \ref{sec:conclusions}).
The importance of repeatability as a basic requirement for
high-quality random number generation has been stressed by L’Ecuyer
\cite{lecuyer2015rng,lecuyer2017parallel}.

Randompack is intended for scientific and statistical computing rather
than cryptographic use. It emphasizes high-throughput generation for
Monte Carlo simulation, bootstrap resampling, stochastic optimization,
and related large-scale computational tasks in which both speed and
statistical quality matter. The library further aims to provide a
small stable C interface, portable implementation, thread-safety, few
external dependencies, clear separation between engine and
distribution layers, support for independent substreams and state
serialization, and reproducibility across platforms. It is also meant
to be modular enough to serve as a test bed for adding new engines and
distributions.

The rest of this article is organized as follows. Section 2 discusses
the primary components of Randompack: the interface, the generators,
and the distributions. Section 3 provides details of the algorithms
used to sample the supported probability distributions. Section 4
describes implementation details, including how modern CPU features
are used to accelerate the computations and how reproducibility across
platforms is achieved. Section 5 describes testing and timing, and the
article concludes with a short Discussion.

\section{Library interface and functionality}
% ===========================================
This section describes the main components of Randompack: the user
interface, the implemented engines, the methods for state
initialization and stream control, and the supported distributions.

\subsection{User interface}
% =====================================================================
The primary entry points are the constructor function
\texttt{randompack\_create}, several sampling functions (one for each
supported distribution), and several seeding, setup, and state control
functions. The constructor returns a handle to an opaque structure
representing a random number generator (RNG). This object contains the
underlying engine, its state, and other internal information needed by
the library. Most library functions return a boolean success value.
When an operation fails, the user can retrieve a diagnostic string
from the RNG object with \texttt{randompack\_last\_error}.

The following minimal example creates an RNG with the default engine,
obtained by passing \texttt{0} to \texttt{randompack\_create}. Then five
samples from the standard normal distribution are drawn and printed out.

\begin{lstlisting}[style=tightcode]
#include <stdio.h>
#include "randompack.h"
int main(void) {
  randompack_rng *rng = randompack_create(0);
  double x[5];
  randompack_norm(x, 5, rng);
  for (int i = 0; i < 5; i++) printf("%g\n", x[i]);
  randompack_free(rng);
  return 0;
}
\end{lstlisting}

A specific engine can be selected by name (engine identifier) e.g.
\texttt{randompack\_create(\"ranlux++\")}. The RNG created by the
constructor is automatically randomized but it can also be manually
seeded after creation, for example with \texttt{randompack\_seed(42,
0, rng)}. The second argument is a spawn key (0 means none is
specified; see Section \ref{sec:state-init}).

\subsection{Implemented engines}
% =====================================================================
Randompack implements 11 different PRNGs or engines.
All these date from the last 20 years. The primary inclusion criteria
for the engines are that they should

\begin{enumerate}
\item have period at least $2^{64}$,
\item have a compact state size,
\item pass rigorous statistical tests, including BigCrush,
\item be computationally efficient, and
\item support the construction of independent substreams.
\end{enumerate}

\newcommand{\scrambledandweb}{\cite{blackman2021scrambled,vigna2026shootout}}
\newcommand{\oneillpcgandpost}{\cite{oneill2014pcg,oneill2019pcg64dxsm-post}}

\begin{table}
\caption{Random number generators (engines) implemented in Randompack.
Three engines have SIMD-accelerated versions that are not listed,
xoshiro256++, xoshiro256**, and sfc64, with identifiers
\texttt{x256++simd}, \texttt{x256**simd}, and \texttt{sfc64simd}, the
first of which has been selected as Randompack's default engine. State
words gives the size of the RNG state in 64-bit words.}
\newcommand{\tsup}[1]{\textsuperscript{#1}}
\label{tab:engines}
\begin{tabular}{llcclcc}
\toprule
Engine & Identifier & State words & Period\tsup{a} & Authors & Year & References \\
\midrule
xoshiro256++   & x256++   & 4 & $2^{256}-1$          & Vigna and Blackman  & 2019 & \scrambledandweb\\
xoshiro256**   & x256**   & 4 & $2^{256}-1$          & Vigna and Blackman  & 2018 & \scrambledandweb\\
xorshift128+   & x128+    & 2 & $2^{128}-1$          & Vigna               & 2014 & \cite{vigna2016experimental}\\
xoroshiro128++ & xoro++   & 2 & $2^{128}-1$          & Vigna and Blackman  & 2019 & \scrambledandweb\\
PCG64 DXSM     & pcg64    & 4 & $2^{128}$            & O'Neill             & 2019 & \oneillpcgandpost\\
squares64      & squares  & 2 & $2^{64}$             & Widynski            & 2021 & \cite{widynski2020squares}\\
Philox-4x64    & philox   & 6 & $2^{256}$            & Salmon and Moraes   & 2011 & \cite{salmon2011parallel}\\
sfc64          & sfc64    & 4 & $\geq 2^{64}$        & Chris Doty-Humphrey & 2010 & \cite{dotyhumphrey2010pract}\\
cwg128         & cwg128   & 8 & $2^{128}$            & Działa              & 2022 & \cite{dziala2023collatz}\\
ranlux++       & ranlux++ & 9 & $\approx 2^{570}$    & Sibidanov           & 2017 & \cite{sibidanov2017ranluxpp}\\
ChaCha20       & chacha20 & 6 & $2^{35}$             & Bernstein           & 2008 & \cite{bernstein2008chacha}\\
\bottomrule
\end{tabular}
\par\medskip
\centering
\begin{minipage}{\dimexpr\linewidth-2cm\relax}
  \footnotesize \raggedright \tsup{a}For PCG64 DXSM and cwg128 the
  listed period is for each increment (out of $2^{127}$), for Philox
  and squares64 it is for each key (out of $2^{128}$ and $2^{64}$,
  respectively, and for ChaCha20 it is for each nonce (out of
  $2^{96}$). For sfc64 the average period is ``a substantial fraction
  of the state space'' \cite{dotyhumphrey2010pract}.
\end{minipage}
\end{table}
Table \ref{tab:engines} lists the selected engines. Most of them have
already been mentioned in the Introduction. The main exceptions are
squares64, a recent fast counter-based RNG, and cwg128, a recent
simple and stream-friendly generator. A few of the generators in the
table do not fulfill all the inclusion criteria: xorshift128+ does not
reliably pass BigCrush, but it is very efficient, does appear in Table
\ref{tab:language-rngs}, and is important in the evolution of the xor
family. It is also useful to include one failing engine to allow
testing Randompack's optional TestU01 driver (see Section
\ref{sec:statistical-quality}). ChaCha20 is not particularly efficient
and its period is relatively short, but it is included to provide one
cryptographically secure engine.

While many of the engines from Table \ref{tab:language-rngs} are
implemented in Randompack, there are some that are absent. Among those
is the most popular one, the Mersenne Twister. Its state is 312 64-bit
words, 35 times more than the largest Randompack engine (ranlux++).
Moreover it fails BigCrush. Of the remaining generators, XORWOW does not
pass BigCrush reliably, but three others, Threefry, Java's LXM
generator, and MRG32k3a, are stronger candidates for later inclusion.
They are high-quality but probably somewhat slower than most of those
present.

With one exception, the C code for the engines is copied directly from
their original or standard public description, with minor adaptation
to the Randompack setting. The exception is ranlux++, a modernized
version of ranlux due to Sibidanov (2017). It was rewritten from his
mathematical description because all available implementations,
including Sibidanov's, are GNU-licensed
\cite{sibidanov2017ranluxpp,hahnfeld2021ranluxpp,jirka2021portable}.

\subsection{State initialization and stream control}
\label{sec:state-init}
%=====================================================================
Randompack provides four classes of RNG engine initialization:
deterministic seeding, explicit stream selection, randomization from
system entropy, and exact assignment of the full internal state.

Seeding deserves careful treatment as some RNGs may show poor
statistical behaviour if seeded directly with a small seed. A better
approach is to mix the seed and spread it across the full state, using
a process that is radically different from the RNG being seeded
\cite{matsumoto2007initialization,oneill2014pcg,blackman2021scrambled}.
As noted in the Introduction, SplitMix \cite{steele2014splitmix} could
be used for this purpose, but O'Neill's \texttt{seed\_seq\_fe128}
\cite{oneill2015seedseq} provides an alternative adopted in
Randompack. It hashes and mixes a vector of seeds into a 128-bit
\emph{entropy pool} and then applies different hash functions to the
pool to generate the output state. O'Neill aims for strict avalanche,
meaning that flipping a bit in the seed material flips each output bit
with probability $1/2$, and reports empirical results consistent with
such behaviour. NumPy's seeding is likewise based on O'Neill's
\texttt{seed\_seq} work.

Randompack’s randomize function obtains (cryptographically secure)
system entropy from the operating system: it uses
\texttt{RtlGenRandom} on Windows, \texttt{arc4random\_buf} on macOS
and OpenBSD, and \texttt{getrandom} on Linux and other Unixes which
support it, with \texttt{/dev/urandom} as a fallback. If that too is
unavailable, an emergency fallback mixes in the system clock.

In Randompack the user seed together with a spawn key of arbitrary
length constitutes the input to the seeding mixer. This creates one
way of creating an arbitrary number of deterministically seeded
substreams. The spawn key could, for instance, consist of MPI rank,
thread id, task number, etc. For Randompack's engines the probability
that two such substreams overlap will be vanishingly small, but, if
the user prefers, Randompack also supports the built-in features of
the different engines for creating mathematically proven independent
streams. Jumps can be used for the xor-family and ranlux, and special
stream setters for the other engines (Table \ref{tab:stream-select}).
Randompack allows jumping by $2^k$ steps, with $k \in
\{32,64,96,128,192\}$ (the last two are not available for
\texttt{x128++} and \texttt{xoro++}). For example, repeatedly jumping
a stream by $2^{64}$ creates several disjoint substreams of length
$2^{64}$. PCG64 is special in that in addition to setting its odd
increment explicitly, it also allows jumping, or \emph{advancing}, its
state by an arbitrary number of steps, $\delta$, with
\texttt{randompack\_pcg64\_advance} (or with \texttt{randompack\_jump}
when $\delta = 2^k$).

\begin{table}
  \caption{State composition and stream selection for engines with
    split state. The actual stream setter functions include the prefix
    \texttt{randompack\_}, e.g. \texttt{randompack\_pcg64\_set\_inc}.
    The setters for cwg128 and sfc64 also apply a mixing warmup after
    setting the stream selector: cwg128 advances by 96 states and
    sfc64 by 18, in line with published guidance on stream
    initialization for the respective generators.}
  \label{tab:stream-select}
  \begin{tabular}{llll}
    \toprule
    Engine   & State layout (64-bit words in parentheses)        & Stream selector  & Stream setter      \\
    \midrule
    pcg64    & state (2) + increment (2)                         & increment        & \texttt{pcg64\_set\_inc}      \\
    squares  & counter (1) + key (1)                             & key              & \texttt{squares\_set\_key}    \\
    philox   & counter (4) + key (2)                             & key              & \texttt{philox\_set\_key}     \\
    sfc64    & state $(a,b,c)$ (3) + counter (1)                 & $(a,b,c)$        & \texttt{sfc64\_set\_abc}      \\
    cwg128   & state (6) + Weyl increment (2)                    & Weyl increment   & \texttt{cwg128\_set\_weyl}    \\
    chacha20 & key (4) + nonce ($1\tfrac{1}{2}$) + counter ($\tfrac{1}{2}$) & nonce & \texttt{chacha20\_set\_nonce} \\
    \bottomrule
  \end{tabular}
\end{table}
Such stream selection can be combined naturally with seed spreading in
parallel simulation. First a base generator is created and seeded.
Then, for subprocess $i$, the generator is duplicated (with the
provided \texttt{randompack\_duplicate} function) and either jumped or
changed with a stream setter. In schematic form:
\begin{lstlisting}[style=tightcode,language=,basicstyle=\ttfamily\small]
create and seed rng[0]
for each subprocess i=1,2,3,...:
  rng[i] = duplicate(rng[i-1])
  set the stream of rng[i] to i (or jump rng[i])
  simulate with rng[i] within subprocess i
\end{lstlisting}

The fourth way of initializing an RNG mentioned at the beginning of
this subsection, exact assignment of the full state, may be done
with \texttt{randompack\_set\_state}.

\begin{table}[t]
\centering
\caption{Randompack sampling functions returning double values for
  continuous distributions. The actual function names include the
  prefix \texttt{randompack\_}, e.g. \texttt{randompack\_u01}. Float
  versions of all the functions except \texttt{mvn} are also provided
  with the suffix \texttt{f}, e.g. \texttt{randompack\_u01f}.}
\label{tab:continuous}
\begin{tabular}{lll}
  \toprule
  Function             & Notation                             & Algorithm \\
  \midrule
  \texttt{u01}         & $U(0,1)$                             & integer-to-double map                                                        \\
  \texttt{unif}        & $U(a,b)$                             & shift-scale transform                                                        \\
  \texttt{norm}        & $N(0,1)$                             & ziggurat (Marsaglia-Tsang, McFarland modified)                               \\
  \texttt{normal}      & $N(\mu,\sigma^2)$                    & shift-scale transform                                                        \\
  \texttt{exp}         & $\mathrm{Exp}(\beta)$                & ziggurat (Marsaglia-Tsang, McFarland modified)                               \\
  \texttt{lognormal}   & $\log N(\mu,\sigma^2)$               & exponential transform                                                        \\
  \texttt{gamma}       & $\Gamma(\alpha,\theta)$              & Marsaglia--Tsang rejection method                                            \\
  \texttt{beta}        & $\mathrm{Beta}(a,b)$                 & $X/(X+Y)$ for independent gamma draws $X,Y$                                  \\
  \texttt{chi2}        & $\chi^2_\nu$                         & gamma special case, $\alpha=\nu/2, \theta=2$                                 \\
  \texttt{t}           & $t_\nu$            & normal--gamma mixture, $Z/\sqrt{V/\nu}$ where $Z\sim N(0,1)$ and $V\sim \chi_\nu^2$            \\
  \texttt{f}           & $F_{\nu_1,\nu_2}$  & gamma ratio, ($X/\nu_1)/(Y/\nu_2)$ where $X\sim\Gamma(\nu_1/2,1)$ and $Y\sim\Gamma(\nu_2/2,1)$ \\
  \texttt{gumbel}      & $\mathrm{Gumbel}(\mu,\beta)$         & inverse CDF transform                                                        \\
  \texttt{pareto}      & $\mathrm{Pareto}(x_m,\alpha)$        & $x_m\exp(E/\alpha)$ where $E$ is $\mathrm{Exp}(1)$                           \\
  \texttt{weibull}     & $\mathrm{Weibull}(k,\lambda)$        & $\lambda \sqrt[k]{E}$ where $E$ is $\mathrm{Exp}(1)$                         \\
  \texttt{skew\_normal}& $\mathrm{SN}(\mu,\sigma,\alpha)$     & Azzalini construction from two independent normal draws                      \\
  \texttt{mvn}         & $N(\mu,\Sigma)$                      & Cholesky factor transform and normal draws                                   \\
  \bottomrule
\end{tabular}
\end{table}

\begin{table}[t]
\centering
\caption{Randompack sampling functions for discrete distributions. The actual function names include the prefix
  \texttt{randompack\_}, e.g. \texttt{randompack\_int}. When the \texttt{uint$n$} functions are called with
  $b = 0$ they generate unbounded integers over the full range $\{0,\ldots,2^n-1\}$.}
\label{tab:discrete}
\begin{tabular}{lll}
  \toprule
  Function            & Notation                          & Description                                \\
  \midrule
  \texttt{int}        & $\mathrm{Unif}\{m,\ldots,n\}$     & 32 bit integers on an interval, unbiased   \\
  \texttt{long\_long} & $\mathrm{Unif}\{m,\ldots,n\}$     & 64-bit integers on an interval, unbiased   \\
  \texttt{uint8}      & $\mathrm{Unif}\{0,\ldots,b-1\}$   & bounded/unbounded unsigned 8-bit integers  \\
  \texttt{uint16}     & $\mathrm{Unif}\{0,\ldots,b-1\}$   & bounded/unbounded unsigned 16-bit integers \\
  \texttt{uint32}     & $\mathrm{Unif}\{0,\ldots,b-1\}$   & bounded/unbounded unsigned 32-bit integers \\
  \texttt{uint64}     & $\mathrm{Unif}\{0,\ldots,b-1\}$   & bounded/unbounded unsigned 64-bit integers \\
  \texttt{perm}       & permutation of $\{0,\ldots,n-1\}$ & Fisher--Yates shuffle                      \\
  \texttt{sample}     & $k$-subset of $\{0,\ldots,n-1\}$  & Floyd$/$reservoir sampling                 \\
  \texttt{raw}        & bitstream                         & raw unbounded bytes into \texttt{void} buffer \\
  \bottomrule
\end{tabular}
\end{table}

\subsection{Distributions}
% ========================
Randompack supports sampling from 14 continuous distributions as
well as a few basic discrete distributions (Tables
\ref{tab:continuous}, \ref{tab:discrete}). Uniformly distributed
samples in $[0,1)$ are provided by \texttt{randompack\_u01} and
general intervals are supported by \texttt{randompack\_unif}. There
are also two functions for the normal distribution,
\texttt{randompack\_norm} for standard normals and
\texttt{randompack\_normal} for general normals, but the other
distributions all have one function each.

Almost all sampling functions have the form
\texttt{randompack\_<dist>} and take parameters:
\begin{samepage}
\begin{enumerate}
\item an array (or buffer) to be filled with generated samples,
\item the number of values to generate,
\item the parameters of the distribution (if any),
\item a handle to an RNG created with \texttt{randompack\_create}.
\end{enumerate}
\end{samepage}
The only exception is \texttt{randompack\_mvn}, for multivariate
normal sampling, that has a more complex parameter list. The function
names and distribution parameters are given in Tables
\ref{tab:continuous} and \ref{tab:discrete}. Use e.g.
\texttt{randompack\_normal(x, 100, 3, 5, rng)} to fill \texttt{x} with
100 $N(3,5)$ samples and \texttt{randompack\_int(100,1,6)} to simulate
100 throws of a die.

More details of the individual distributions are given in the next
section, which also discusses the underlying algorithms.

\section{Algorithms for distributions}

\subsection{Continuous distributions}
\label{sec:cont-distr}
% =====================================================================
\textbf{Uniform distribution.} The randompack engines all compute a
stream of bits that can be interpreted as either 64-bit or 32-bit
words. In the following $U(0,1)$-doubles are considered, the float
case being analogous. Let $u$ be a 64-bit word, interpreted as an
integer in $[0,2^{64})$. Two common methods exist to convert $u$ to a
double $x$ in $[0, 1)$:
\begin{samepage}
\begin{enumerate}
\item Let $x = \left\lfloor u/2^{11} \right\rfloor 2^{-53}$.
\item Put $\lfloor u/2^{12}\rfloor$ in the fraction field of $x$, and 1023
in its exponent field, giving $x \in [1,2)$. Then subtract 1.
\end{enumerate}
\end{samepage}
The first method corresponds to \texttt{x = (u >> 11) * 0x1.0p-53} and
gives a 53-bit resolution floating-point number. The second method,
giving 52-bit resolution, often compiles to faster code and is the one
used by default in Randompack. In C it may be obtained with:
\begin{lstlisting}[style=tightcode]
bits = (u >> 12) | 0x3ff0000000000000ULL;
memcpy(&x, &bits, 8);
x -= 1;
\end{lstlisting}
Randompack offers a full-mantissa function to select between the two
resolutions.

\textbf{Normal and exponential distributions.} The ziggurat method
since 2000 \cite{marsaglia2000ziggurat,doornik2005improved}, discussed
above, has become quite widespread in recent software: NumPy, Julia,
Matlab, Octave, GSL, Boost, Rust, Go, and Java all implement some
version of it. Randompack uses the same geometry as most of these
libraries with 256 equal-area strips, all but the base strip
rectangular. Following NumPy, each normal and exponential draw begins
with drawing a 64-bit word. Its low byte is used to select the strip
number; for normals bit 8 is used to select the sign of a candidate
sampled $x$ and bits 9--60 determine its magnitude; for exponentials
bits 8--60 determine the magnitude of $x$. In both cases the high 3
bits are unused.
\begin{figure}[t]
\centering
\includegraphics[width=0.7\textwidth]{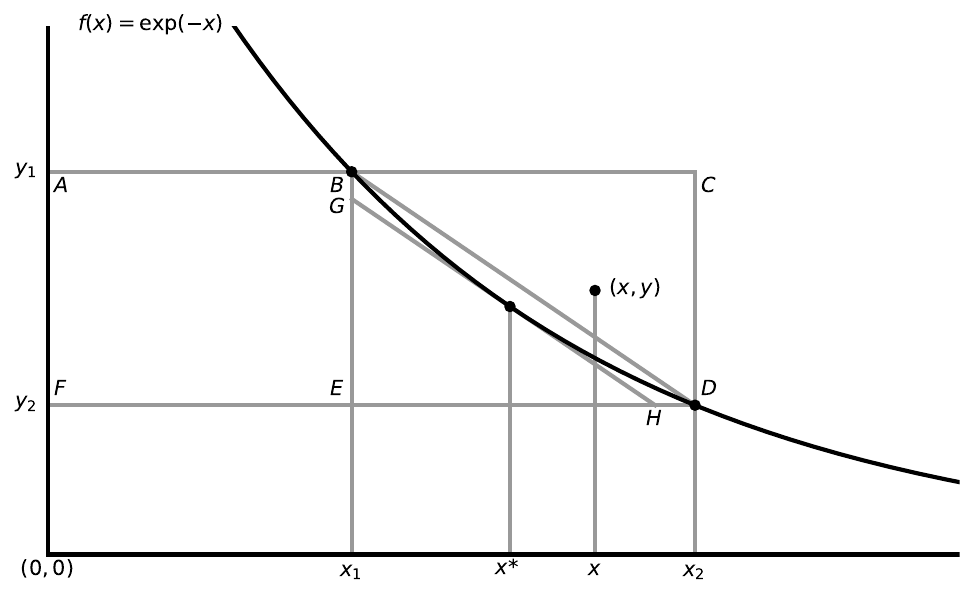}
\caption{Illustration of the ziggurat method for the exponential
  distribution and 3 strips: Strip 1 is $[0,x_1] \times [y_1,1]$,
  strip 2 is the rectangle $ACDF$, and the base strip is the region
  below $FD$ together with the tail. Initially a random strip is
  selected (here strip 2) along with an $x$ within its limits ($x <
  x_2$). With $l(x)$ denoting the $y$-value on the secant $BD$, $x^*$
  maximizes $d_2(x) := |f(x) - l(x)|$ and $GH$ is the tangent at
  $(x^*, f(x^*))$. Also, let $d_2^* = d_2(x^*)$. Initially $x \in [0,
  x_2)$ is drawn. If $x < x_1$ (the fast accept region) the algorithm
  returns $x$ immediately. If not, the original 2000-version next
  draws $y \in [y_2, y_1)$, computes $f(x)$ immediately, accepts $x$
  if $y < f(x)$ and rejects it otherwise. Randompack also draws $y$ in
  the same interval, accepts $x$ if $y$ is below the tangent and
  rejects it if $y$ is above the secant, and only computes $f(x)$ if
  $(x,y)$ is in the trapezium $BDHG$. With McFarland's method strip 2
  is the trapezium $ABDF$. If $x$ is not fast-accepted, a fresh
  $(x,y)$ is drawn in $\triangle BDE$. One could now accept if $y$ is
  below the tangent, but, to save on precomputed constants, McFarland
  actually suggests accepting if $y < l(x) - \max_i d_i$. If $y$ is
  too big it is compared with a freshly computed $f(x)$.}
\Description{Diagram of the ziggurat method for exponential sampling,
  showing three horizontal strips under a decreasing density curve,
  labeled points defining rectangles, secants, a tangent, acceptance
  regions, and the tail region used by the algorithm.}
\label{fig:ziggurat}
\end{figure}
The 2000 ziggurat algorithm has a 0.069\% rejection rate for normal
sampling and 1.10\% for exponential sampling (Figure
~\ref{fig:ziggurat}). It requires 15 PDF-evaluations (and thus exp
evaluations) for 1000 normal samples and 23 for 1000 exponential
samples. In 2016 McFarland \cite{McFarland2016} proposed an improved
version, where most of these exp evaluations are saved. It has been
adopted by the Java standard library. The method employs trapezoidal
strips with much lower rejection rate. Figure ~\ref{fig:ziggurat}
illustrates an interval with convex PDF; for the concave part of the
normal PDF the situation is similar, except that there the tangent is
above the secant, and in the interval containing $x=1$ (the inflection
point of the normal PDF) there are two tangents.

The triangle sampling of \cite{McFarland2016} is somewhat costly, and
therefore Randompack uses the original strip definitions but
incorporates McFarland-type savings in PDF computations (Figure
\ref{fig:ziggurat}). For each 1000 normal draws it requires only 1.2
exp evaluations and for the exponential distribution the rate is half
of that. For efficiency, as in \cite{marsaglia2000ziggurat}, the
implementation translates the test $x < x_{i-1}$ to an integer test
involving precomputed constants $k_i$, and similarly the comparison of
$y$ with $l(x)$ and $l(x) \pm d_i^*$ is translated to an integer test
with precomputed integer versions of $d_i^*$. The $k_i$, as well as
double constants $w_i = 2^{-52}x_{i-1}$ and $f_i = y_i$ also described
in \cite{marsaglia2000ziggurat} are copied verbatim from the NumPy
source code, but the $d_i^*$ have been computed by solving equations
$f'(x) = s$. The solution is $x^* = \log s$ for the exponential
distribution and $x^* = \sqrt{-W_k(-s^2)}$ for the normal
distribution, where $W_k$ is the $k$-th branch of the Lambert $W$
function, $k=0$ for $x^* < 1$ and $k=-1$ for $x^* > 1$.

The resulting savings in execution time, compared with always
computing $f(x)$ when $x$ is not fast-accepted, have been measured as
7\%–18\% for normal draws and 15\%–23\% for exponential draws
depending on the platform.

\textbf{Gamma and related distributions}. Randompack implements the
gamma sampling of the 2000 article \cite{marsaglia2000gamma}. Like the
ziggurat method dating from the same year it has been widely adopted:
NumPy, Matlab, GSL, R rangen, TensorFlow, and Julia all use it. A good
gamma generator is important as four distributions implemented in
Randompack rely on it, cf. Table ~\ref{tab:continuous}. These are
implemented using the formulae given in the table, which can all be
found in modern textbooks on statistical theory.

\textbf{Extreme value and Weibull distributions.} The Gumbel
distribution sampling is implemented by supplying U(0,1) samples to
its inverse CDF, $F^{-1}(u)=\mu-\beta \log(-\log u)$. Pareto
distribution samples are obtained via transformation of exponential
samples (Table \ref{tab:continuous}). To draw samples $X$ from a
generalized Pareto distribution with parameters $(\mu, \sigma, \xi)$
where $\xi \neq 0$, let $P$ be $\mathrm{Pareto}(1, 1/|\xi|)$ and
compute
$$
X =
\begin{cases}
  \mu + \sigma(P-1)/\xi   & \xi > 0 \\
  \mu - \sigma(1-1/P)/\xi & \xi < 0
\end{cases}
$$
If $\xi = 0$ then $X$ is a shifted exponential distribution, $X \sim
\mu + \mathrm{Exp}(\sigma)$. The Weibull distribution has widespread
application e.g. in survival analysis, wind speed modelling, and
hydrology to name a few. Randompack samples it as the scaled $k$-th
root of an $\mathrm{Exp}(1)$ distribution as shown in the table.

\textbf{Multivariate normal distribution.} Contrary to all the other
continuous distributions, Randompack does not support a single
precision version of \texttt{randompack\_mvn}. To generate a $d\times
n$ matrix $X$ with $n$ replicates of $d$-variate normal samples with
covariance $\Sigma$, a Cholesky factorization $\Sigma = LL^T$ is first
attempted. If that fails (due to $\Sigma$ being indefinite) a pivoted
Cholesky factorization $\Sigma=AA^T$ (where $A$ need not be lower
triangular) with the Lapack \emph{dpstrf} is tried. If either
factorization succeeds, a $d \times n$ array $Z$ of independent
$N(0,1)$ variates is now drawn and $X$ is computed as either $\mu +
LZ$ or $\mu + AZ$. For convenience randompack also allows $X$ to be
$n\times d$.

\subsection{Discrete distributions}
% =====================================================================
\textbf{Uniform bounded integers.} Randompack provides functions for
sampling various integer data types from specified intervals (Table
\ref{tab:discrete}). On almost all current systems int and long long
correspond to 32-bit and 64-bit integers, respectively, and Randompack
only supports such systems. With no loss of generality sampling from
$\{0,1,\ldots,b-1\}$ can be considered. When $b$ is not a power of 2
rejection sampling is needed to get unbiased samples, and Randompack
implements the efficient 2019 rejection method of Lemire
\cite{lemire2019fastrandom}, which replaces divisions used by many
earlier methods with much cheaper multiplications. Languages and
libraries that have adopted this method include NumPy, Julia, Rust
rand, Go, GNU and Microsoft C++ standard libraries, and C\# on .NET.

\textbf{Permutations and sampling.} As many other random libraries,
Randompack implements the Fisher--Yates shuffle
\cite{fisher1938statistical,durstenfeld1964algorithm235}, that
computes an $n$-element permutation by repeatedly choosing a random
not yet chosen integer less than $n$.

Sampling a set $S$ of $k$ elements from $\{0,1,\ldots,n-1\}$ is a
little more complex. When $k$ is small compared to $n$ Floyd's
algorithm \cite{knuth1969art} is efficient: Begin with
$S=\varnothing$; for $j=n-k$ to $n-1$ draw a number from $1,\ldots,j$;
if it is not in $S$ add it; otherwise add $j$ to $S$. This is
efficient if the set is implemented with a simple hash map. When $k$
is bigger there are several options e.g. a partial Fisher--Yates
shuffle, but Randompack uses a reservoir sampling method which avoids
memory allocation when $k>n/2$ \cite{vitter1985reservoir}.

\textbf{Unbounded integers and raw bitstreams}. As shown in Table
\ref{tab:discrete}, Randompack supports unbounded integers by setting
the b argument to the \texttt{uint}$n$ functions to 0. In addition
\texttt{randompack\_raw} can be used to fill a buffer of any type with
random bits.

\section{Implementation details}

\subsection{SIMD acceleration}
% =====================================================================
Randompack makes substantial use of vector SIMD instructions on modern
CPUs. There are three main SIMD architectures in use: NEON (on ARM
processors), and AVX2 and AVX-512 (on X86-64 processors). The provided
SIMD instructions allow operating in parallel on 32 bit or 64 bit
words, with two 64-bit lanes for NEON, four for AVX2 and eight for
AVX-512. In addition, modern CPUs can often execute several
instruction chains in parallel through instruction-level parallelism
(ILP). Both SIMD and ILP depend on the computations being independent
and can be realized through loop unrolling and multiple parallel state
updates.

In Randompack this CPU-level parallelism is used in four main places.
First, three engines have SIMD implementations, xoshiro256++,
xoshiro256**, and sfc64 (Table \ref{tab:bench-engines}). Second, SIMD
instructions are used in shift--scale transformations, e.g. to get
$U(a,b)$ from $U(0,1)$, and also to get general normal samples from
standard normal samples. Third, they are used in the integer-to-double
conversion functions. Fourth, vectorized logarithms and exponentials
from the Sleef library \cite{sleef_github,shibata2020sleef} are used
to speed up the sampling of all the supported continuous distributions
except uniform, exponential, and normal.

Together, SIMD and ILP yield substantial speed-ups, with the SIMD
engines providing the largest part of the gain. To achieve the
necessary independent computations, eight disjoint scalar streams are
created for each engine. For xoshiro256++ and xoshiro256** they are
obtained by jumping a base engine by the values $k2^{253},
k=1,\ldots,7$ and for sfc64 they are obtained by adding $k2^{61},
k=1,\ldots,7$ to the engine's counter, with wraparound if the sums
exceed $2^{64}$. In practice, this arrangement does not interfere with
the jumping and stream selection functions discussed in Section
\ref{sec:state-init}. Repeatedly jumping x256++simd by $2^{192}$
creates disjoint sequences of length $2^{61}$. With Randompack on a
fast computer, it would take more than 20 years to exhaust one such
sequence. The sequences obtained by setting sfc64simd with different
$(a,b,c)$ triplets also have minimum length $2^{61}$.

\subsection{Exact reproducibility across platforms}
% =====================================================================
Nearly all modern computers use the same floating-point representation
for C floats and doubles. This is set by the IEEE 754 standard, which,
in addition to the format, specifies rounding rules and mandates that
the results of the basic arithmetic operations +, –, *, and /, as well
as square roots and fused multiply-add (fma;
$\operatorname{fma}(a,b,c)=a*b+c$), shall produce results correctly
rounded. This holds for Intel and AMD X86-64 as well as ARM and RISC-V
systems. It also carries over to other programming languages that
offer 32-bit and 64-bit floating point numbers, such as C++, Fortran,
Python, Julia and R.

Randompack's raw bitstreams, being computed by integer and bit
operations, are inherently platform-agnostic, and the same holds for
doubles and floats derived from them by multiplication and addition.
As Randompack uses the same algorithms on all platforms, this means
that the discrete and $U(0,1)$ distributions are all bit-identical
across platforms. However, the IEEE 754 standard does not set in stone
the computation of trigonometric functions, logarithms, and
exponentials. To overcome this, Randompack uses the OpenLibm library
\cite{openlibm,gladman2025accuracy} to compute logarithms and
exponentials in its rejection tests in the ziggurat and gamma
functions. This library is designed to work consistently across
compilers and operating systems, and when used with compiler options
that disable optimizations that can change the result of
floating-point calculations (e.g. \texttt{-ffp-contract=off},
\texttt{-ffp-model=precise} for clang) it gives bit-identical results
on Apple, Windows, and several Linux systems with gcc, clang, icx, and
MSVC. This ensures that the bit-identity is extended to the $N(0,1)$,
$\mathrm{Exp}(1)$ and $Gamma(\alpha, 1)$ with $\alpha \geq 1$. Apart
from multivariate normal (mvn) the remaining distributions depend
continuously on the underlying random bits, and therefore the
generated samples agree to within a few ulp even if non-bit-identical,
but faster, Sleef logarithms and exponentials are used to compute
them.

Randompack offers a setting, \texttt{bitexact}, that makes all
distributions except mvn bit-identical across platforms by using
OpenLibm instead of Sleef for all library function calls, and by using
scalar rather than SIMD code for affine transformations such as
converting $U(0,1)$ to $U(a,b)$. Turning on \texttt{bitexact} makes
generation of lognormal, gamma with shape $< 1$, Gumbel, Pareto, and
Weibull samples 2--4 times slower (depending on platform). For the
remaining continuous distributions that use affine transformations,
\texttt{bitexact} slows generation by a few percent.

\subsection{Buffered architecture}
% ================================
A small buffer is used between the engine and distribution layers.
When the buffer is empty, the selected engine refills it with raw
random bits; the distribution functions then consume these buffered
words as \texttt{uint64\_t} or \texttt{uint32\_t} values, either singly
or in bulk. This arrangement is important to the structure of the
library, since it allows many engines and many distributions to
coexist cleanly in one implementation without engine-specific
distribution code. The buffer size must be a multiple of 72 64-bit
words, as ranlux++ fills it in 9 word chunks and the SIMD-engines as
well as chacha20 fill it in 8 word chunks. The current buffer size is
144 words, a value selected by approximate benchmarking experiments to
give good performance.

\subsection{Miscellaneous features}
% =====================================================================
\textbf{Thread safety and checkpointing.} Randompack is thread-safe
provided that each thread uses its own RNG object. There is no global
state, and error conditions belong to individual RNGs. The package is
accompanied by several examples demonstrating the use of parallel
random number generation. 

A serialization function allows checkpointing long-running simulation
jobs. It encodes the current state of an RNG in an opaque byte buffer.
This buffer may be used to restore the state exactly at a later time.
Jobs can then be restarted if they stop prematurely for some
reason. This is particularly relevant when massively parallel jobs are
run, with their associated increased likelihood of failures.

\textbf{Engine query.} The function \texttt{randompack\_engines}
reports the supported engine identifiers together with short
descriptions that include the engines' traditional full names, their
authors and publication years, and the state size. The interface
deliberately uses a single canonical identifier for each engine; it is
case-insensitive, but apart from that no aliases exist.

\textbf{Float versions.} Almost all the continuous Randompack
distributions exist in both double precision (C \texttt{double}) and
single precision (C \texttt{float}) versions. The float functions have
an \texttt{f} appended to their name, e.g. \texttt{randompack\_expf}.
For many simulation applications there is no need for the extra
precision of doubles, so both space and time can be saved by using
floats.

\textbf{Jumping} Let $s$ be an $n$-bit state and $T$ the state
transition of an xor-family engine. Jumping $s$ by $N$ steps amounts
to computing $T^Ns$. Since $T$ is linear on an $n$-dimensional vector
space over $F_2$, $T^N = p(T)$ for an $F_2$-polynomial of degree $< n$ (a
consequence of the Cayley-Hamilton theorem). $T^N$ can be obtained
with repeated squaring and the coefficients of the jump polynomial $p$
by Gaussian elimination. The Python program \texttt{jumpgeneral.py},
which accompanies Randompack, implements this method, and it was used
to compute the xor-family jump polynomials used in Randompack. The
method of \cite{haramoto2008efficient} could also have been employed,
but that method is a little more involved; however it is applicable
for larger $n$, such as that of the Mersenne Twister. The
\texttt{ranlux++} jumps are computed with
\texttt{compute\_ranlux\_jumps.c} also supplied with Randompack,
whereas the PCG64 advancing, described in \cite{oneill2014pcg}, is
copied from NumPy. The correctness of the computed jump polynomials is
systematically checked by the test suite.

\section{Validation and performance}
% ==================================
The quality of a random-number library has several aspects: the
implementation should be correct, the algorithms should be efficient,
and the generated streams should satisfy stringent statistical tests.

This section discusses how Randompack meets these requirements, in
some cases using a belt-and-suspenders approach because errors in
random number generation can have serious consequences while being
hard to infer from simulation output alone.

\subsection{Test suite} 
% =====================
\label{subsec:testing}
Randompack includes a comprehensive test suite which has been run on
five platforms (Table \ref{table:platforms}). The tests fall into
three categories:
\begin{enumerate}
\item Tests of correct API behaviour
\item Validation of sample series, internal and external
\item Checking of statistical quality
\end{enumerate}
The API testing includes testing expected error reporting, testing
edge cases (e.g. zero and maximum lengths should work, clean failure
on null buffer and rng pointers), correct default engine behaviour and
rejection of incorrect engine names, etc. The basic functionality of
all user-facing functions is exercised through one or more calls. In
short, this testing covers aspects of the API other than the
correctness of the drawn samples.

\begin{table}
  \caption{Systems used for testing and benchmarking Randompack. The
    tests were run with clang on all systems and gcc on all but
    Windows. Intel icx was also used on Windows and the workstation,
    nvc on DGX-Spark, and MSVC on Windows. All tests passed.}
\label{table:platforms}
\centering
\begin{tabular}{llll}
  \toprule
  Computer          & CPU                      & SIMD support & Operating system \\
  \midrule
  Linux workstation & Intel Core i5-1335U      & AVX2         & Ubuntu           \\
  Linux server      & Intel Xeon Platinum 8358 & AVX-512      & Rocky Linux      \\
  DGX-Spark         & Arm Cortex-X925          & NEON         & Ubuntu           \\
  Apple MacBook     & Apple M4 (Arm)           & NEON         & macOS            \\
  Windows           & Intel Core i7-14700KF    & AVX2         & Windows 11       \\
  \bottomrule
\end{tabular}
\end{table}

The internal sample validation checks that same seeds give same sample
streams and different seeds give different streams. It checks that the
SIMD-engines agree with their scalar counterparts. It also checks that
serialization followed by deserialization gives an RNG that matches
the original one, and the same applies to a duplicated RNG. The
external validation compares the following engines' raw output:
\begin{enumerate}
\item pcg64 against NumPy,
\item chacha20 against the RFC8439 published test vector,
\item philox against Random123 reference output,
\item x256++ against Rust,
\item x256** against Rust,
\item ranlux++ against \cite{jirka2021portable} and \cite{hahnfeld2021ranluxpp}.
\end{enumerate}
The actual test programs contain more details of these comparisons in
comments. In addition the tests compare the output of
\texttt{seed\_seq\_fe128}, which Randompack uses for seeding, with the
original \cite{oneill2015seedseq}, bit for bit (see Section
\ref{sec:state-init}). Also, the OpenLibm logarithm and exponential
functions are compared with the system library counterparts.

Several tests assess the statistical quality of sampled sequences. The
tests in the suite look at the draws in isolation, ignoring possible
correlation between successive draws which, however, is covered by the
BigCrush tests discussed below. Uniform sequences are checked for
histogram balance via binning, and they are transformed to normal with
the probit function and the mean, variance, skewness and kurtosis
checked against theory. There is also a min/max check based on the
order statistics of $U(0,1)$. For all the other continuous
distributions probability integral transform (PIT) is used to
transform the sequences to be uniform, and the results are tested in
the same way.

The discrete distributions are compared similarly for support (no
draws outside specified result intervals), balanced counts of each
possible outcome, bit balance for draws in $[0,2^k)$, and
combinatorial validity for \texttt{randompack\_perm} and
\texttt{randompack\_sample}.

\subsection{Statistical quality}
% ==============================
\label{sec:statistical-quality}
Randompack supports running the TestU01 batteries, but they must be
installed separately for licensing reasons. The engines selected for
Randompack all pass the BigCrush tests of TestU01 when their raw bit
output is piped directly into BigCrush in bit-stream mode, according
to the original publications describing the engines. This was also
confirmed in the author's own tests. However, when the bits are
reversed before feeding them to BigCrush, one engine fails,
\texttt{xorshift128+}, confirming previously published results
\cite{lemire2019xorshiftfails}. Since the ziggurat algorithm is partly
new, separate probability integral transform (PIT) tests
\cite{diebold1998evaluating} of normal and exponential samples from
Randompack were also carried out by feeding the transformed samples to
BigCrush in U(0,1) mode, with fail-free results. Both the raw-bit
tests and the PIT tests were done by checking that in 100 BigCrush
runs the total numbers of p-values below 0.001, 0.0001, and 0.00001
were no greater than expected by chance.

\begin{table}[t]
\centering
\caption{Engine throughput timings in nanoseconds per 64 output bits.
  The table shows averages over 10 runs of 1-second timings on five
  computers (see Table \ref{table:platforms}), with the benchmark
  programs compiled using clang}
\label{tab:bench-engines}
\setlength{\tabcolsep}{4pt}
\begin{tabular}{
  ll
  S[table-format=1.2]
  S[table-format=1.2]
  S[table-format=2.2]
  S[table-format=1.2]
  S[table-format=2.2]
}
\toprule
Engine & Identifier & {Mac-M4}
& {Spark} & {Xeon}
& {Core-i5} & {Win} \\
\midrule
xoshiro256++, SIMD accelerated & \texttt{x256++simd} & 0.30 & 0.41 &  0.30 & 0.44 &  0.26 \\
xoshiro256**, SIMD accelerated & \texttt{x256**simd} & 0.40 & 0.54 &  0.33 & 0.47 &  0.29 \\
sfc64, SIMD accelerated        & \texttt{sfc64simd}  & 0.32 & 0.34 &  0.30 & 0.33 &  0.22 \\
xoshiro256++                   & \texttt{x256++}     & 0.73 & 0.81 &  1.07 & 0.88 &  0.77 \\
xoshiro256**                   & \texttt{x256**}     & 0.76 & 0.80 &  1.15 & 0.83 &  0.79 \\
xorshift128+                   & \texttt{x128+}      & 0.76 & 0.88 &  1.14 & 0.91 &  0.78 \\
xoroshiro128++                 & \texttt{xoro++}     & 0.99 & 0.82 &  1.36 & 1.16 &  0.96 \\
PCG64-DXSM                     & \texttt{pcg64}      & 1.17 & 1.56 &  1.73 & 1.26 &  1.02 \\
sfc64                          & \texttt{sfc64}      & 0.72 & 0.70 &  1.04 & 0.91 &  0.54 \\
squares64                      & \texttt{squares}    & 0.82 & 0.99 &  1.99 & 1.25 &  0.97 \\
Philox-4x64                    & \texttt{philox}     & 1.07 & 1.77 &  2.75 & 1.51 & 10.46 \\
cwg128                         & \texttt{cwg128}     & 0.91 & 1.25 &  1.55 & 1.22 &  0.97 \\
ranlux++                       & \texttt{ranlux++}   & 2.69 & 4.23 &  8.12 & 4.79 &  6.08 \\
ChaCha20                       & \texttt{chacha20}   & 7.39 & 8.20 & 13.10 & 9.76 & 10.01 \\
\bottomrule
\end{tabular}
\end{table}

\begin{table}[t]
  \centering
  \caption{Benchmark timings for continuous distributions. The table
    shows average nanoseconds per value over 10 runs of 0.5-second
    timings, obtained by repeatedly filling a buffer with 4096 values.
    See Table \ref{table:platforms} for details of the computers. The
    first four columns show timings for the C program but benchmarking
    in C++, Julia, and Fortran gave essentially identical results.
    These results are based on clang compilation; gcc was also tested
    and was 1\%–16\% slower across the platforms, while on the
    workstation icx was 6\% slower than clang. }
  \label{tab:bench-distributions}
  \setlength{\tabcolsep}{2pt}
  \setlength{\cmidrulewidth}{\lightrulewidth}
  \begin{tabular}{
      l
      S[table-format=1.2]
      S[table-format=2.2]
      S[table-format=2.2]
      S[table-format=1.2]
      S[table-format=2.2]
      c@{\hspace{1em}}
      S[table-format=1.1]
      S[table-format=2.1]
      c@{\hspace{1em}}
      S[table-format=1.1]
      S[table-format=2.1]
      c@{\hspace{1em}}
      S[table-format=2.1]
      S[table-format=2.1]
      S[table-format=2.1]
      S[table-format=2.1]
      S[table-format=1.1]
    }
    \toprule
    & \multicolumn{11}{c}{R a n d o m p a c k} & \multicolumn{6}{c}{} \\
    \cmidrule(r){2-12}
    & \multicolumn{5}{c}{C, C++, Julia, Fortran} && \multicolumn{2}{c}{Python} && \multicolumn{2}{c}{R}
    && \multicolumn{5}{c}{Other libraries (on Core-i5)} \\
    \cmidrule(r){2-6}
    \cmidrule(r){8-9}
    \cmidrule(r){11-12}
    \cmidrule{14-18}
    Distribution & {Spark} & {Xeon} & {Win} & {Mac} & {Core-i5}
    && {Mac} & {Core-i5} && {Mac} & {Core-i5}
    && {NumPy} & {Base-R} & {Julia} & {C++} & {MKL} \\
    \midrule
    %%              –––––––––––––––––––––R a n d o m p a c k ––––––––––––––––––––––––    –––––––– Other libraries ––––––
    %% .out files:   spark &  xeon &windows&  Mac &   i5  &&pymac& pyi5 &&R-mac& R-i5 && NumPy&   R  & Julia&  cpp & MKL
    $U(0,1)$        & 0.44 &  0.29 &  0.30 & 0.34 &  0.42 && 0.4 &  0.5 && 2.1 &  1.8 &&  1.8 &  6.7 &  0.7 &  6.3 & 0.5 \\
    $U(2,5)$        & 0.47 &  0.30 &  0.31 & 0.34 &  0.41 && 0.5 &  0.5 && 2.1 &  1.7 &&  2.7 &  6.7 &  0.8 &  6.4 & 0.5 \\
    Std. normal     & 1.2  &  1.8  &  1.2  & 0.9  &  1.4  && 1.0 &  1.6 && 2.8 &  2.7 &&  8.9 & 21.4 &  1.4 & 14.3 & 2.3 \\
    Normal$(2,3)$   & 1.3  &  2.0  &  1.3  & 1.0  &  1.5  && 1.1 &  1.7 && 2.7 &  2.7 &&  9.6 & 21.4 &  1.4 & 14.3 & 2.3 \\
    Exponential     & 1.2  &  1.7  &  1.1  & 1.0  &  1.3  && 1.1 &  1.5 && 2.7 &  2.7 &&  4.2 & 23.5 &  1.4 & 10.6 & 1.6 \\
    Log-normal      & 2.6  &  2.9  &  2.0  & 1.8  &  2.5  && 1.9 &  2.7 && 7.1 &  3.8 && 13.9 & 28.7 &  4.0 & 18.9 & 3.0 \\
    Skew-normal     & 2.6  &  4.5  &  3.9  & 2.1  &  3.2  && 2.2 &  3.5 && 3.7 &  5.0 && 27.9 &  \NA &  8.2 & \NA  & \NA \\
    Gumbel          & 4.8  &  3.2  & 11.6  & 2.6  &  3.7  && 2.6 &  3.8 && 8.3 &  5.5 && 13.3 &  \NA &  8.4 & \NA  & \NA \\
    Pareto          & 2.9  &  3.4  &  8.7  & 2.1  &  2.9  && 2.1 &  3.1 && 4.3 &  4.7 && 17.5 &  \NA &  4.4 & \NA  & \NA \\
    Gamma$(2,3)$    & 3.9  &  9.7  &  8.2  & 3.6  &  5.4  && 3.6 &  5.6 && 5.6 &  7.0 && 15.7 & 53.4 &  8.8 & 27.5 & 6.1 \\
    Beta$(2,5)$     & 8.1  & 19.7  & 17.4  & 7.3  & 11.0  && 7.2 & 11.4 && 9.2 & 13.6 && 31.3 & 56.2 & 20.0 & \NA  & 8.9 \\
    Chi-square$(5)$ & 3.8  &  9.1  &  7.9  & 3.5  &  5.3  && 3.5 &  5.5 && 5.3 &  7.0 && 15.7 & 53.4 &  8.7 & 27.2 & 6.0 \\
    $t(10)$         & 5.8  & 15.5  & 16.9  & 4.8  &  9.6  && 4.8 &  9.7 && 6.4 & 11.1 && 25.3 & 60.2 & 15.4 & \NA  & \NA \\
    $F(5,10)$       & 7.9  & 19.2  & 17.3  & 7.2  & 11.0  && 7.0 & 11.1 && 9.0 & 13.4 && 32.5 & 98.2 & 25.7 & \NA  & \NA \\
    Weibull$(3,4)$  & 4.9  &  4.3  & 12.4  & 3.0  &  4.2  && 3.1 &  4.3 && 9.9 &  5.7 && 14.1 & 24.2 & 31.4 & 23.1 & 2.9 \\
    \bottomrule
  \end{tabular}
\end{table}

\begin{table}
  \caption{Benchmark timings for discrete distributions. The table
    shows average nanoseconds per value over 10 runs of 0.5 second
    timings. The bounded-integer results were obtained by repeatedly
    filling a buffer with 4096 values, while the other rows were
    obtained by repeatedly creating a permutation or drawing a sample
    without replacement of the indicated size. Results are for the
    Linux workstation (Table \ref{table:platforms}), with clang used
    for the compilation of Randompack. C++, Julia, and Fortran times
    were essentially identical to the C times.}
  \label{tab:bench-integers}
  \setlength{\tabcolsep}{3pt}
  \centering
  \begin{tabular}{
      l
      l
      S[table-format=1.1]
      S[table-format=3.1]
      S[table-format=3.1]
      c@{\quad}
      S[table-format=3.1]
      S[table-format=3.1]
      S[table-format=3.1]
      S[table-format=2.1]
      S[table-format=1.1]
    }
    \toprule
    && \multicolumn{3}{c}{Randompack} && \multicolumn{5}{c}{Other libraries} \\
    \cmidrule(lr){3-5} \cmidrule(){7-11}
    {Distribution} & {Parameters} & {C} & {Python} & {R} && {NumPy} & {Base-R} & {Julia} & {C++} & {MKL} \\
    \midrule
    bounded-integer & $1,10$                  & 0.7 &  0.8 &   1.9 &&   3.1 &  28.4 &   2.2 &  2.3 & 0.6 \\
    bounded-integer & $1,10^5$                & 0.7 &  0.8 &   1.6 &&   3.0 &  27.4 &   2.2 &  2.2 & 0.6 \\
    bounded-integer & $1,2\!\cdot\!\!10^9$    & 1.3 &  1.4 &   2.3 &&   7.7 &  20.1 &   2.2 &  2.2 & 0.6 \\
    bounded-integer & $1,6\!\cdot\!\!10^{18}$ & 1.1 &  1.3 &   \NA &&   6.4 &   \NA &   4.8 &  5.9 & \NA \\
    permutation     & $100$                   & 3.8 & 11.5 &  29.3 &&  17.9 &  59.7 &   3.9 &  1.6 & \NA \\
    permutation     & $100000$                & 3.6 &  3.8 &   4.3 &&   8.3 &  30.2 &   4.0 &  1.7 & \NA \\
    sample          & $20,1000$               & 9.7 & 51.4 & 139.6 && 203.7 & 205.9 & 201.9 & 81.9 & \NA \\
    sample          & $500,1000$              & 9.9 & 16.1 &  16.1 &&  19.5 &  30.9 &   8.6 &  3.2 & \NA \\
    sample          & $980,1000$              & 3.5 &  4.4 &   6.7 &&  13.7 &  27.0 &   4.5 &  1.7 & \NA \\
    \bottomrule
  \end{tabular}
\end{table}

\subsection{Performance}
% ======================
Several benchmarking programs accompany Randompack, as detailed in the
readme file. These benchmark both continuous and discrete sampling
using the C library, as well as the Fortran, Python, R, and Julia
interfaces mentioned in the Introduction. The present article does not
describe those interfaces further, but some of the benchmark tables
include their performance in order to illustrate the practical use of
Randompack from those environments. There are also timing programs for
several other random number generation libraries including the built-in
generators of Python, R, and Julia.

The performance of the implemented engines on the platforms in Table
\ref{table:platforms} is reported in Table \ref{tab:bench-engines}.
Tables \ref{tab:bench-distributions} and \ref{tab:bench-integers}
report benchmarking of the different continuous and integer
distributions on selected platforms. These two tables also include
comparison with some other libraries on the Linux workstation. We note
that Randompack is often several times faster than these libraries.
For $U(0,1)$ it is 3, 3.7, 1.7, 15, and 1.3 times faster than Python,
R, Julia, C++, and MKL, respectively, for $N(2,3)$ the factors are
approximately 6, 8, 1, 10, and 2, and for integers in
$\{1,2,\ldots,10\}$ they are ca. 4, 14, 3, 3, and 1. For Python and R
this comparison applies to the respective language interfaces of
Randompack (in order to compare apples to apples).

\section{Conclusions and availability}
\label{sec:conclusions}
Randompack provides a portable C implementation of modern
pseudo-random number generators and distribution samplers. Its
generators are selected for statistical robustness, speed, and
independent substream mechanisms. The design cleanly separates engines
and distributions, allowing any engine to be used with any
distribution. Fast engines, SIMD acceleration, and modern sampling
algorithms make Randompack faster overall than the competing libraries
shown in Tables \ref{tab:bench-distributions} and
\ref{tab:bench-integers}, sometimes many times faster.

The default engine, the SIMD-version of xoshiro256++
(\texttt{x256++simd}) was chosen over the ** variety because it is
slightly faster and better suited for SIMD vectorization. While
\texttt{sfc64simd} is often faster still, distinct settings of its
$(a,b,c)$ state lack the provable guarantee of disjoint substreams
that holds for jumping the xor-family generators.

A further contribution of the current work is the inclusion of a
permissively licensed implementation of ranlux++, which to the author's
knowledge is the first such implementation available. The Randompack
version is also 2--3 times faster than the program of
\cite{hahnfeld2021ranluxpp} and as fast or up to twice as fast as the
(non-portable) program of \cite{sibidanov2017ranluxpp} (data not
shown).

The source code of Randompack, including benchmarks, test suite, and
examples is publicly available on the project's GitHub page
\texttt{https://github.com/jonasson2/randompack}, where a
comprehensive readme-file may be found. Binary builds of the library
for most modern platforms are also available on
\texttt{https://conda-forge.org} and installable with \texttt{conda
  install -c conda-forge randompacklib}. An important feature of
Randompack is reproducibility across platforms and across interfaced
programming languages. For $U(0,1)$, $N(0,1)$, $\textrm{Exp}(1)$, and
all the discrete distributions the agreement is bit-identical. For other
distributions except multivariate normal the default mode is
tolerance-reproducible, with agreement to within a few ulp, and a
bitexact mode can be selected with an option.

As mentioned in the Introduction interfaces to the Randompack C
library from other languages---Fortran, R, Python and Julia---have
been created, and in addition it can be used directly from C++, as
\texttt{randompack.h} contains an \texttt{extern "C"} guard. The
source code of the interfaces is also available on the GitHub page,
and binary versions are downloadable from the respective languages'
usual distribution platforms, PyPI and conda-forge for Python, CRAN
for R, and the Julia General Registry for Julia. The Fortran interface
is included with the C interface.

\begin{acks}
The author used ChatGPT and OpenAI Codex as writing and programming
assistants for copy editing, wording suggestions, code review, and
debugging support. The author reviewed and is responsible for all text,
code, results, and references in this work.
\end{acks}

%%
%% Bibliography
%%
\bibliography{toms}

%%
%% Appendix (optional)
%%

\end{document}